\documentclass[showpacs,twocolumn,aps,prl]{revtex4-2}
\usepackage{graphicx}
\usepackage{dcolumn}
\usepackage{bm}
\usepackage{amsmath}
\usepackage{amssymb}
\usepackage[usenames,dvipsnames]{xcolor}
\usepackage{color}
\usepackage[colorlinks,plainpages=false]{hyperref}

\begin{document}

\title{Strange quark matter and proto-strange stars in an equivaparticle model}
\author{H.~M.~Chen,$^1$ C.~J.~Xia,$^2$ G.~X.~Peng,$^{1,3,4}$}

\affiliation{%
$^1$\mbox{School of Nuclear Science and Technology, University of Chinese Academy of Sciences, Beijing 100049, China}\\
$^2$\mbox{School of Information Science and Engineering, Ningbo Institute of Technology, Zhejiang University, Ningbo 315100, China}\\
$^3$\mbox{Theoretical Physics Center for Science Facilities, Institute of High Energy Physics, P.O. Box 918, Beijing 100049, China}\\
$^4$\mbox{Synergetic Innovation Center for Quantum Effects and Application, Hunan Normal University, Changsha 410081, China}  }

\begin{abstract}
The properties of strange quark matter and the structures of (proto-)strange stars are studied within the framework of an equivparticle model, where a new quark mass scaling and self-consistent thermodynamic treatment are adopted. Our results show that the perturbative interaction has a strong impact on the properties of strange quark matter. It is found that the energy per baryon increases with temperature, while the free energy decreases and eventually becomes negative. At fixed temperatures, the pressure at the minimum free energy per baryon is zero, suggesting that the thermodynamic self-consistency is preserved. Additionally, the sound velocity $v$ in quark matter approaches to the extreme relativistic limit ($c/\sqrt{3}$) as the density increases. By increasing the strengths of confinement parameter $D$ and perturbation parameter $C$, the tendency for $v$ to approach the extreme relativistic limit at high density is slightly weakened. For (proto-)strange stars, in contrast to the quark mass scalings adopted in previous publications, the new quark mass scaling can accommodate massive proto-strange stars with their maximum mass surpassing twice the solar mass at  $T = 50$ MeV.
\end{abstract}


\maketitle

\section{Introduction}
\label{intro}

Since the energy per baryon decreases as one increases the degrees of freedom, in 1984 Witten speculated that the 3-flavor strange quark matter (SQM) might be more stable than the 2-flavor nuclear matter, i.e., the true ground state of strong interactions \cite{Witten1984}. It was shown that SQM play important roles in many interesting fields. For example, SQM can accommodate many abnormal astronomical phenomena in both neutron stars \cite{Bombaci2006} and strange stars \cite{Perez-Garcia2010}. In particular, the presence of SQM may be essential to understand the $r$-mode instability of compact stars \cite{XuJF2021CPC} in the the multi-messenger era of astronomy \cite{LIGO2017PRL}. The droplets made of SQM, namely strangelets or slets~\cite{Alford1999,Rajagopal2001,Madsen2001,Peng2006}, could serve as a unique signature for the formation of quark gluon plasma (QGP) in the experiments of relativistic heavy-ion collision. A through investigation on the properties of SQM is thus crucial for us to understand various astrophysical phenomena as well as the QCD phase transition mechanisms \cite{Peng2008,Wen2013,Burgio2002,Logoteta2013,Orsaria2014}.

In principle, the properties of quark matter can be described accurately adopting quantum chromodynamics (QCD). However, because of the non-perturbative nature at small energy densities and the sign problem in lattice simulations, QCD can not be solved from the first principle. Effective models reflecting QCD characteristics are widely used to study the stability and properties of SQM, e.g., the Nambu-Jona-Lasinio model \cite{Nambu1961}, perturbation model \cite{Fraga2005}, field correlator method \cite{Plumari2013}, quark-cluster model \cite{Shi2003,R.2010} and other models \cite{Khadekar2011,Wen2013a,Isayev2013,Wang2010,Huang2004,Bao2008,Peng2000,Peng1999,Wen2005,Hou2015}. Most of the effective models adopt quark masses that vary with density (chemical potential) and/or temperature~\cite{Goloviznin1993, Peshier1994, Gorenstein1995, Bluhm2005, Bannur2007, Gardim2009, Li2010, Luo2013, Wu2005, Lu2016a}, where we refer to those with density-dependent quark masses as equivparticle model \cite{Chakrabarty1989, Benvenuto1995, Lugones1995, Peng2000}.

The purpose of this work is to investigating the properties of SQM at finite temperature and density, where an equivparticle model including both confinement and first-order perturbation interactions is adopted. The paper is organized as follows. In Sec.~\ref{inconThermo}, we discuss self-consistent thermodynamic treatment.
In Sec.~\ref{sec:therm}, we present the quark mass scale for strange quarks at finite temperatures with both confinement and perturbation interaction in mind.
In Sec.~\ref{sec:SQM}, we give numerical results for the properties of strange quark matter.
In Sec.~\ref{stars}, we give numerical results for the properties of strange stars.
A summary is given in Sec.~\ref{sec:sum}.

\section{\label{inconThermo} Self-consistent thermodynamic treatment}
One important aspect for equivparticle model is the thermodynamic self-consistency, where the basic thermodynamic relations need to be fulfilled, e.g., the pressure at the minimum energy/free-energy per baryon should be exactly zero. A detailed investigation on thermodynamically self-consistent treatments was carried out in Ref.~\cite{Xia2014}, where the quark masses depend on the density and/or temperature \cite{PhysRevC.91.015208}. The basic thermodynamic differential relation of free energy $\overline{F}$ reads
\begin{equation}\label{1}
\mathrm{d}\overline{F}=-\overline{S}\mathrm{d}T-P\mathrm{d}V+\sum_{i}\mu_{i}\mathrm{d}{N_{i}},
\end{equation}
where $\overline{S}$, $T$, $P$, $V$, $\mu_{i}$, and $N_{i}$ represents the the entropy, temperature, pressure, volume, chemical potential and particle number of the system, respectively. For a uniform system, one can define the free energy density $F =\overline{F}/V $, entropy density $S =\overline{S}/V$, and particle number density $n_{i} =N_{i}/V$. Substitute those into Eq.~(\ref{1}), we have
\begin{eqnarray}\label{dF}
\mathrm{d}F
&=&-S\mathrm{d}T+\left(-P-F+\sum_{i}\mu_{i}n_{i}\right)\frac{\mathrm{d}V}{V} \nonumber\\
& &{}+\sum_{i}\mu_{i}\mathrm{d}n_{i}.
\end{eqnarray}
For a system comprised of noninteracting particles, the free energy density is connected to the thermodynamic potential density $\Omega_{0}(T,V,\{\mu_{i}\},\{m_{i0}\})$ by
\begin{equation}
F = \Omega_{0} + \sum_{i}\mu_{i}n_{i}. \label{Eq:F}
\end{equation}
Note that the masses $m_{i0}$ take constant values in those equations. To account for the strong interaction among quarks and gluons, the masses are replaced with the equivalent ones $m_{i}=m_{i}(n_{u},n_{d},n_{s},T)=m_{i0}+m_\mathrm{I}(n_{u},n_{d},n_{s},T)$, where $m_{i0}$ is the current mass of quarks. In such cases, as the mass varies with the state variables $T$ and $n_{i}$, the entropy density, pressure, and chemical potential are altered~\cite{PhysRevC.91.015208, Xia2014}. Here we consider the temperature $T$, volume $V$, and quark number densities $n_{i}$ as independent state variables, while the free energy density $F$ corresponds to the characteristic thermodynamic function. The free energy density takes the same form as the free particle system with the quark masses replaced by the equivalent ones in Eq.~(\ref{Eq:F}), i.e.,
\begin{eqnarray}\label{F}
F &=& F(T,V,\{n_{i}\},\{m_{i}\})  \nonumber\\
  &=& \Omega_{0}(T,V,\{\mu_{i}^{*}\},\{m_{i}\})+\sum_{i}\mu_{i}^{*}n_{i},
\end{eqnarray}
where the chemical potential $\mu_i$ is replace with the effective one $\mu_i^*$. The differential relation then becomes
\begin{eqnarray}\label{ddF}
\mathrm{d}F &=& \frac{\partial\Omega_{0}}{\partial T}\mathrm{d}T+\frac{\partial\Omega_{0}}{\partial V}\mathrm{d}V\nonumber\\
          & &{} +\sum_{i}\left(\frac{\partial\Omega_{0}}{\partial \mu_{i}^{*}}\mathrm{d}\mu_{i}^{*}+\mu_{i}^{*}\mathrm{d}n_{i}+n_{i}\mathrm{d}\mu_{i}^{*}\right)\nonumber\\
          & &{} +\sum_{i}\frac{\partial\Omega_{0}}{\partial m_{i}}\left(\sum_{j}\frac{\partial m_{i}}{\partial n_{j}}\mathrm{d}n_{j}+\frac{\partial m_{i}}{\partial T}\mathrm{d}T\right) \nonumber\\
          &=& \left[
      \frac{\partial \Omega_0}{\partial T}
      + \sum_i \frac{\partial \Omega_0}{\partial m_i} \frac{\partial m_i}{\partial T}
\right]\mbox{d}T
+ \frac{\partial \Omega_0}{\partial V} \mbox{d}V
\nonumber \\
&&
+
\sum_i \left[ \mu_i^* +  \sum_j\frac{\partial \Omega_0}{\partial m_j}\frac{\partial m_j}{\partial n_i}
\right] \mbox{d} n_i.
\end{eqnarray}
Comparing Eq.~(\ref{dF}) with Eq.~(\ref{ddF}), one immediately obtains the thermodynamic quantities
\begin{eqnarray}
S &=& -\frac{\partial\Omega_{0}}{\partial T}-\sum_{i}\frac{\partial m_{i}}{\partial T}\frac{\partial\Omega_{0}}{\partial m_{i}}, \label{7} \\
P &=&  -F+\sum_i \mu_{i} n_i - V\frac{\partial \Omega_0}{\partial V}, \label{8} \\
\mu_{i} &=& \mu_i^* +   \sum_j\frac{\partial \Omega_0}{\partial m_j}\frac{\partial m_j}{\partial n_i}. \label{9}
\end{eqnarray}
The particle number densities $n_{i}$ and the energy density are then fixed by
\begin{eqnarray}
n_{i} &=& -\frac{\partial\Omega_{0}}{\partial \mu_{i}^{*}}, \label{10} \\
E &=& F+TS=\Omega_{0}+\sum_{i}\mu_{i}^{*}n_{i}+TS. \label{11}
\end{eqnarray}

\section{\label{sec:therm} Density and/or temperature dependent particle masses}
Another important issue for equivparticle model concerns the quark mass scaling, i.e., how to parameterize the density and/or temperature dependent particle masses. The equivalent quark masses are usually comprised of two parts, i.e.,
\begin{equation}
m_i=m_{i0}+m_\mathrm{I},
\end{equation}
where $m_{i0}$ represents the current mass ($m_{u0}=5\ \mathrm{MeV}$, $m_{d0}=10\ \mathrm{MeV}$, and $m_{s0}=100\ \mathrm{MeV}$) of quarks and $m_\mathrm{I}$ accounts for the strong interaction. Considering the fact that quark confinement dominates at low densities, based on MIT bag model, the quark mass scaling was initially parameterized as inversely proportion to the baryon number density~\cite{Fowler1981, Chakrabarty1989, Chakrabarty1991, Chakrabarty1993}, i.e.,
\begin{equation}
m_\mathrm{I}=\frac{B}{3n_\mathrm{b}}
\end{equation}
where $B$ is the bag constant. The contribution from temperature can be considered in an expansion form \cite{Zhang2002}
\begin{equation}
m_\mathrm{I}=\frac{B}{3n_\mathrm{b}}\left[1-\left(\frac{T}{T_{c}}\right)^{2}\right].
\end{equation}
An inversely cubic scaling was later derived considering the in-medium quark condensate and linear confinement \cite{Peng1999a, Peng2000}, i.e.,
\begin{equation}
m_\mathrm{I} = \frac{D}{n_\mathrm{b}^{1/3}},
\end{equation}
where $D$ is the quantity reflecting the confinement strength. Then this mass scaling was extended to finite temperature scenarios by taking~\cite{Wen2005}
\begin{equation}
m_\mathrm{I} = \frac{D}{n_\mathrm{b}^{1/3}}\left[1-\frac{8T}{\lambda T_{c}}\mathrm{exp}\left(-\frac{\lambda T_{c}}{T}\right)\right], \label{Eq:mI_Wen}
\end{equation}
where $\lambda=1.6$ and $T_{c}=175$ MeV is the critical temperature for the transition of hadronic matter into QGP. This quark mass scaling is consistent with the temperature dependence of the string tension. However, $m_\mathrm{I}$ becomes negative at temperatures above the critical value $T_\mathrm{c}$, while we expect $m_\mathrm{I}\rightarrow 0$ at large temperatures. The inversely cubic scaling can be extended to include the one-gluon-exchange interaction~\cite{Chen2012}. To account for the first-order perturbative interactions at larger densities, an additional term proportional to the cube root of baryon number density was obtained in Ref.~\cite{Xia2014}, i.e.,
\begin{equation}
m_\mathrm{I} = \frac{D}{n_\mathrm{b}^{1/3}}+Cn_\mathrm{b}^{1/3},   \label{Eq:mI_perturb}
\end{equation}
where $C$ represents the strength of perturbative interactions. If $C$ takes negative values, the one-gluon-exchange interaction is effectively included~\cite{Chen2012}. The effects of finite temperature in Eq.~(\ref{Eq:mI_perturb}) was later considered for $ud$ quark matter \cite{Lu2016}, i.e.,
\begin{eqnarray}
m_\mathrm{I} &=& \frac{D}{n_\mathrm{b}^{1/3}}\left(1+\frac{8T}{\Lambda}e^{-{\Lambda}/{T}}\right)^{-1}\nonumber\\
             & & {} +Cn_\mathrm{b}^{1/3}\left(1+\frac{8T}{\Lambda}e^{-{\Lambda}/{T}}\right)   \label{mass}
\end{eqnarray}
with $\Lambda = 280$ MeV. In contrast with the mass scaling in Eq.~(\ref{Eq:mI_Wen}), here the quark masses tends to zero above the critical temperature $T_{c}$. We thus adopt this mass scaling to further investigate the properties of SQM at finite temperatures.

According to our previous study, the parameters lie within the range $120\ \mathrm{MeV} \leq \sqrt D \leq 270\ \mathrm{MeV}$ and $C\leq 1.1676$. In order for SQM to be more stable than nuclear matter and still consistent with traditional nuclear physics, the parameters $D$ and $C$ need to be fixed carefully so that the minimum energy per baryon for SQM is less than 930 MeV while $ud$ quark matter is larger than 930 MeV. In this work, we adopt the parameter sets ($C$, $\sqrt D$ in MeV) as (0.7,129), (0.6,160), (0.4,160), (0,160), (0.4,140), and (0.4,129).

At large temperatures, one needs to consider the contribution of gluons as well. The equivalent gluon mass is defined as
\begin{equation}
(m_{g}/T)^{2}=\eta\alpha\theta(T-T_{c}),
\end{equation}
with $T_{c}=175$ MeV and $\eta = 15${~\cite{Lu2016}}. For the running coupling constant $\alpha$, we use a rapidly convergent expansion \cite{Peng2006a}
\begin{equation}
\alpha=\frac{\beta_{0}}{\beta_{0}^{2}\ln{(u/\Lambda_{T})}+\beta_{1}\ln\ln(u/\Lambda_{T})},
\end{equation}
where $\beta_{0}=11/2-N_{\mathrm{f}}/3$, $\beta_{1}=51/4-(19/12)N_{\mathrm{f}}$, and $N_{\mathrm{f}}=3$ is the number of quark flavors of SQM. To include the non-perturbative effects, we assume that the renormalization scale varies linearly with temperature, i.e.,
\begin{equation}
u/\Lambda_{T}=c_{0}+c_{1}x,
\end{equation}
where $c_{0}=1$, $c_{1}=1/2$, $x=T/T_{c}$.

The density and temperature derivative of quark and gluon masses are essential to obtain the properties of quark matter in equivparticle model, which are obtained with
\begin{eqnarray}
\frac{\partial m_{q}}{\partial n_{b}} &=& -\frac{D}{3n_{b}^{4/3}}\left(1+\frac{8T}{\Lambda}e^{-\Lambda/T}\right)^{-1}\nonumber\\
                                      & &{} +\frac{C}{3n_{b}^{2/3}}\left(1+\frac{8T}{\Lambda}e^{-\Lambda/T}\right), \\
\frac{\partial m_{q}}{\partial T}     &=&  -\frac{8D}{\Lambda n_{b}^{1/3}} \left(1+\frac{\Lambda}{T}\right) \frac{e^{-\Lambda/T}}{\left(1+\frac{8T}{\Lambda}e^{-\Lambda/T}\right)^{2}} \nonumber\\
                                      & &{} +8Cn_{b}^{1/3}\left(\frac{1}{T}+\frac{1}{\Lambda}\right)e^{(-\Lambda/T)}, \label{derivation} \\
\frac{\mathrm{d}m_{g}}{\mathrm{d}T}   &=& \frac{m_{g}}{T}\left(1+\frac{x}{2}\frac{\mathrm{d}\ln \alpha}{\mathrm{d}x}\right),
\end{eqnarray}
and
\begin{equation}
\frac{\mathrm{d}\ln \alpha}{\mathrm{d}x}=-\frac{c_{1}\alpha}{c_{0}+c_{1}x}\left[\beta_{0}+\frac{\beta_{1}}{\beta_{0}}\frac{1}{\ln(c_{0}+c_{1}x)}\right].
\end{equation}

Finally, it is worth mentioning that there are other forms of quark mass scalings in the literature, e.g., an explicit isospin dependent term can be introduced to include quark matter symmetry energy~\cite{Chu2013, Wang2021_Galaxies}, the asymptotic freedom of strong interaction can be considered explicitly through a Wood-Saxon factor \cite{Peng2016}. In this paper, we investigate the properties of SQM at finite temperature adopting the quark mass scaling in Eq.~($\ref{mass}$), where both quark confinement and first-order perturbation interactions are included.

\section{properties of strange quark matter }
\label{sec:SQM}

For simplicity, we assume that the system consists of SQM and gluons at finite temperatures.
According to the equivalent mass model, the contribution of free particles to the thermodynamic potential density can be written as

\begin{equation}
\Omega_{0}=\Omega_{0}^{+}+\Omega_{0}^{-}+\Omega_{0}^{g}.
\end{equation}

The contribution of the equivalent particle (+) and the antiparticle (-) is
\begin{equation}
\Omega_{0}^{\pm}=\sum_{i}-\frac{d_{i}T}{2\pi^{2}}\int_{0}^{\infty}\ln\left[1+e^{-(\sqrt{p^{2}+m_{i}^{2}}\mp\mu_{i}^{*})/T}\right]p^{2}\mathrm{d}p.
\end{equation}

\begin{equation}
\Omega_{0}^{g}=\frac{d_{g}T}{2\pi^{2}}\int_{0}^{\infty}\ln\left[1-e^{-\sqrt{p^{2}+m_{g}^{2}}/T}\right]p^{2}\mathrm{d}p,
\end{equation}
where $i=q$($q=u,d,s$), $d_{q}=3(colors)\times2(spins)=6$ and $i=e$, $d_ {e}=2$ and $i=g$, $d_{g}=8(colors)\times2(spins)=16$.

Substitute this thermodynamic potential into Eqs.~(\ref{7})-(\ref{11}), we obtain the thermodynamic density of each particle at a finite temperature.
The number densities of particle and anti-particle is
\begin{equation}\label{35}
n_{i}^{\pm}=\frac{d_{q}}{2 \pi^{2}}\int_{0}^{\infty}\frac{p^{2}\mathrm{d}p}{e^{(\sqrt{p^{2}+m_{i}^{2}}\mp \mu_{i}^{*})/T}+1},
\end{equation}
The chemical potentials are obtained with Eq.~(\ref{9}), where the derivative terms with respect to equivalent masses are, giving
\begin{eqnarray}
\frac{\partial \Omega_{0}}{\partial m_{q}}&=&\frac{d_{q} m_{q}}{2\pi^{2}}\int_{0}^{\infty}[\frac{1}{e^{(\sqrt{p^{2}+m_{q}^{2}}-\mu_{q}^{*})/T}+1}\nonumber\\
& &+\frac{1}{e^{(\sqrt{p^{2}+m_{q}^{2}}+\mu_{q}^{*})/T}+1}]\frac{p^{2}\mathrm{d}p}{\sqrt{p^{2}+m_{q}^{2}}}.
\end{eqnarray}

\begin{equation}\label{37}
\frac{\partial \Omega_{0}}{\partial m_{g}}=\frac{d_{g} m_{g}}{2\pi^{2}}\int_{0}^{\infty}\frac{1}{e^{\sqrt{p^{2}+m_{g}^{2}}/T}-1}\frac{p^{2}\mathrm{d}p}{\sqrt{p^{2}+m_{q}^{2}}}.
\end{equation}

To obtain the properties of SQM inside compact stars, we introduce electron $e$, the condition of charge neutrality can be fulfilled.
In addition, due to the weak reactions such as  $d$,$ s  \leftrightarrow  u + e + \overline{\nu}_{e} $ and $s + u \leftrightarrow  u + d $, the chemical potential $ \mu_{i}$ ($i = u, d, s, e$) needs to satisfy weak equilibrium conditions (where we have assumed vanishing neutrino chemical potentials for simplicity).
\begin{equation}\label{38}
\mu_{u}^{*}+\mu_{e}^{*}=\mu_{d}^{*}=\mu_{s}^{*},
\end{equation}
and the charge neutrality condition is
\begin{equation}\label{39}
2n_{u}^{+}-n_{d}^{+}-n_{s}^{+}-3n_{e}^{+}-(2n_{u}^{-}-n_{d}^{-}-n_{s}^{-}-3n_{e}^{-})=0,
\end{equation}
and the quark number density satisfies the baryon number conservation condition
\begin{equation}\label{40}
n_{b}=\sum_{q=u,d,s}\frac{1}{3}\left(n_{q}^{+}-n_{q}^{-}\right).
\end{equation}

In the framework of the thermodynamic treatment method given in Sec.~\ref{inconThermo} , we could connect particle number density Eq.~(\ref{35}) with effective chemical potential $\mu_{i}^{*}$ through the particle number density expression .
Therefore, when we have a given baryon number density, the formula of Eqs.~(\ref{38})-(\ref {40}) is comprised of four effective chemical potential equations.
So we can get the effective chemical potential by solving these four equations. Eqs.~(\ref{7})-(\ref{11}) were used to find other thermodynamic quantities.
The real chemical potential $\mu_ {i} $ connect with the effective chemical potential $\mu_ {i} ^ {*} $ by formula Eq.~(\ref {9}).
\begin{figure}
\centering
\includegraphics[width=8cm]{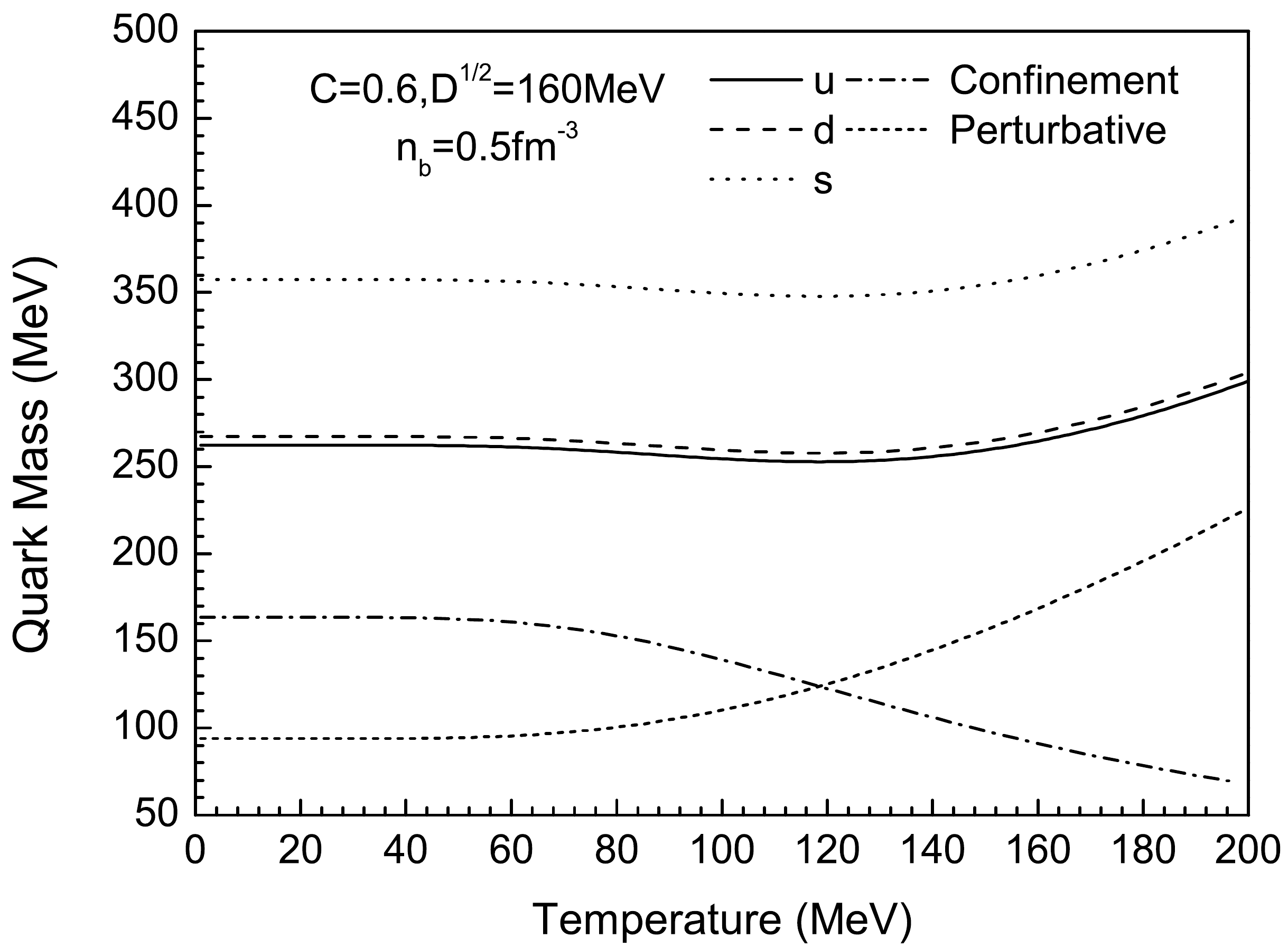}
\caption{ The mass of quark as a function of temperature at $n_{b}=0.5\ \mathrm{fm}^{-3} $, where $D^{1/2}=160\ \mathrm{MeV}$ and $C= 0.6$.}
\label{Fig1}
\end{figure}

According to our quark mass scale, we show a curve of density as a function of temperature when $n_\mathrm{b}= 0.5\ \mathrm{fm}^{-3} $, $D^{1/2}= 160\ \mathrm{MeV}$ and $C=0.6$ in Fig.~\ref{Fig1}.
The contributions of confinement and perturbation are drawn by dashed line and dash-dotted line respectively.
We note that the quark mass constraint becomes less important than the perturbation term as the temperature increases.
\begin{figure}
\centering
\includegraphics[width=8cm]{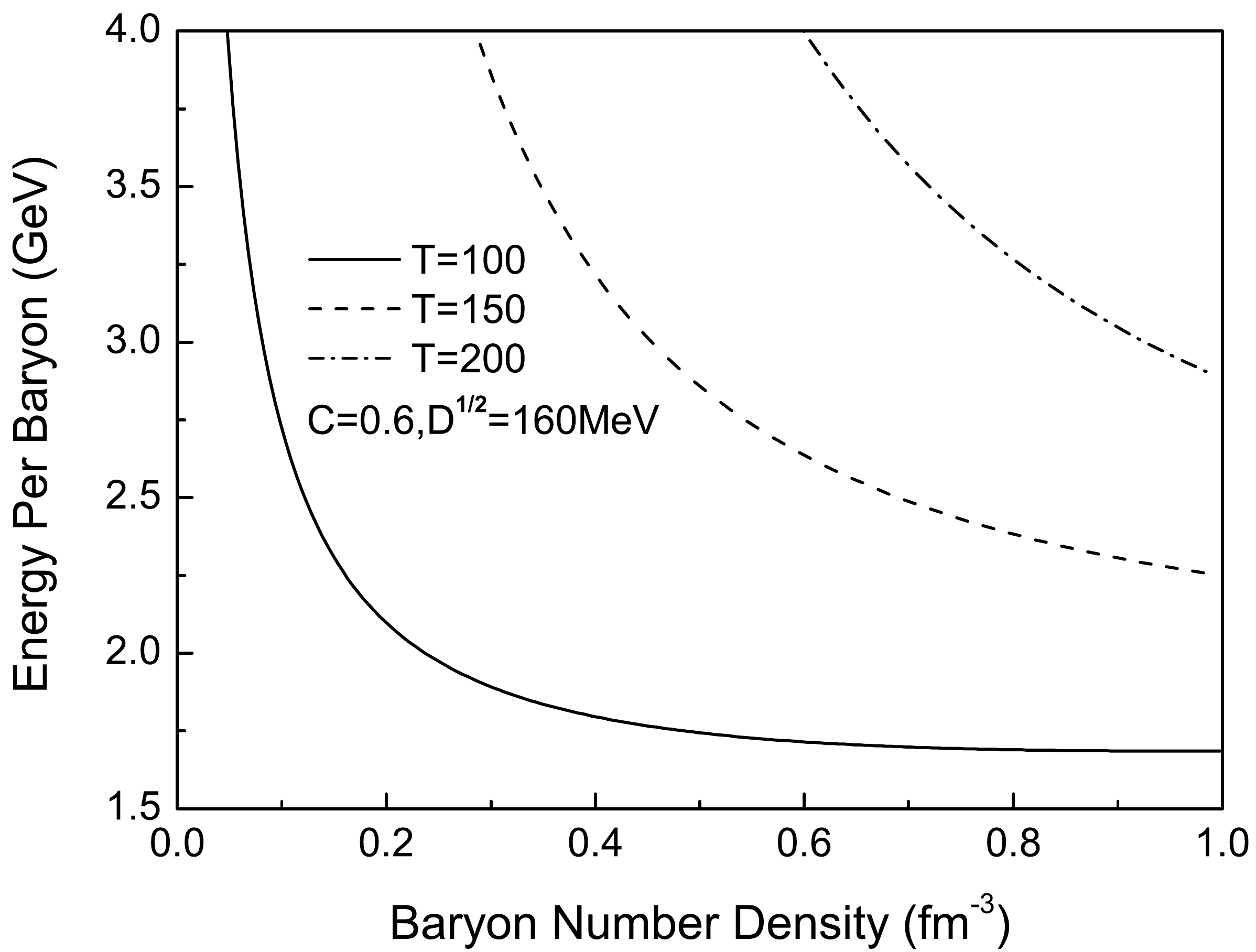}
\caption{Energy per baryon as a function of baryon number density at different temperatures, with $T=100\ \mathrm{MeV}$, $150\ \mathrm{MeV}$, $200\ \mathrm{MeV}$.}
\label{Fig2}
\end{figure}

In Fig.~\ref{Fig2}, we consider the energy per baryon as a function of the baryon number density, where $T=100\ \mathrm{MeV}$, $150\ \mathrm{MeV}$ , $200\ \mathrm{MeV}$ and $D^{1/ 2}=160\ \mathrm{MeV}$ and $C= 0.6$ are adopted. It is found that, at small densities, the energy per baryon becomes infinite, represented by quark confinement.
At a higher baryon density, the energy per baryon also tends to increase because perturbation interactions are becoming more and more important.
\begin{figure}
\centering
\includegraphics[width=8cm]{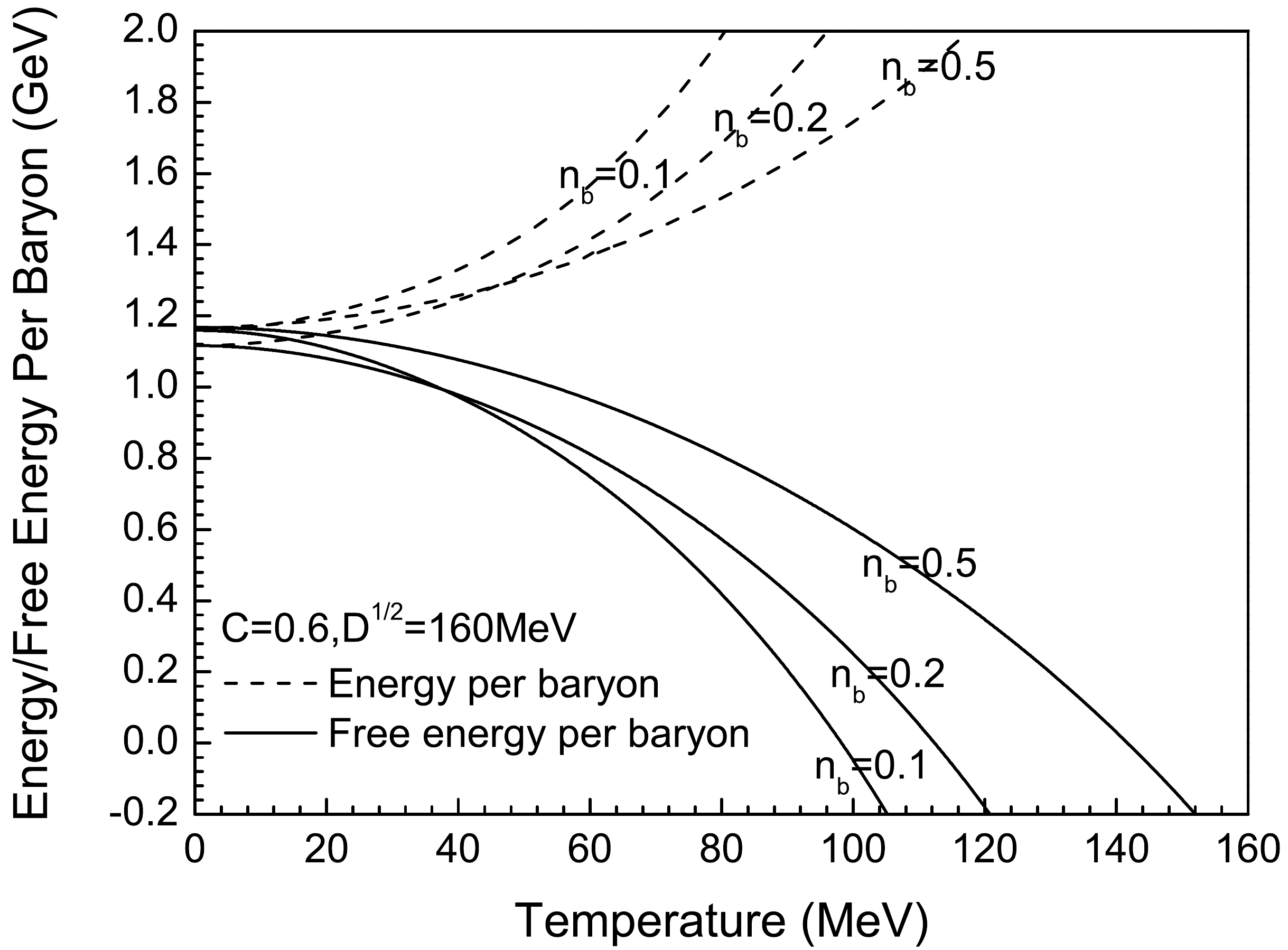}
\caption{The relationship between the free energy per baryon and energy per baryon as functions of temperature at different densities.}
\label{Fig3}
\end{figure}

The temperature dependence of the energy per baryon and the free energy per baryon is shown in Fig.~\ref{Fig3}.
We notice that the energy per baryon increases with temperature.
However, the free energy per baryon decreases with temperature and eventually becomes negative.
Consider the fact that $F=E-TS$ and the entropy term $-TS$ dominate at high temperatures, this is understandable.
\begin{figure}
\centering
\includegraphics[width=8cm]{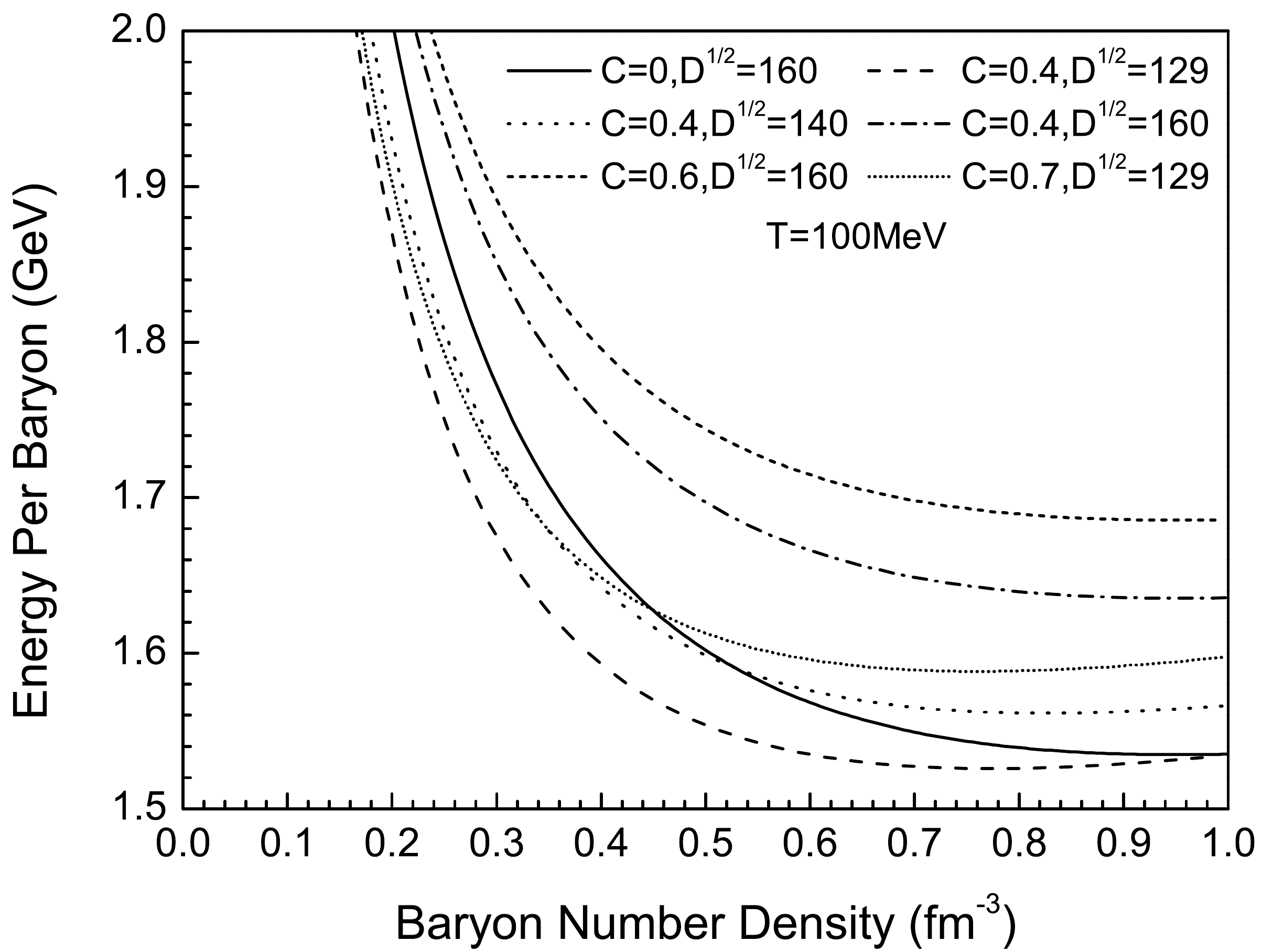}
\caption{The relationship between the energy per baryon of strange quark matter and the baryon number density, where $T=100\ \mathrm{MeV}$.}
\label{Fig4}
\end{figure}

The obtained equation of state(EOS) of matter of SQM inside compact stars is presented in Fig.~\ref{Fig4}.
We find that the slope of EOS increases with the increase of $C$ and decrease of $D$.

\begin{figure}
\centering
\includegraphics[width=8cm]{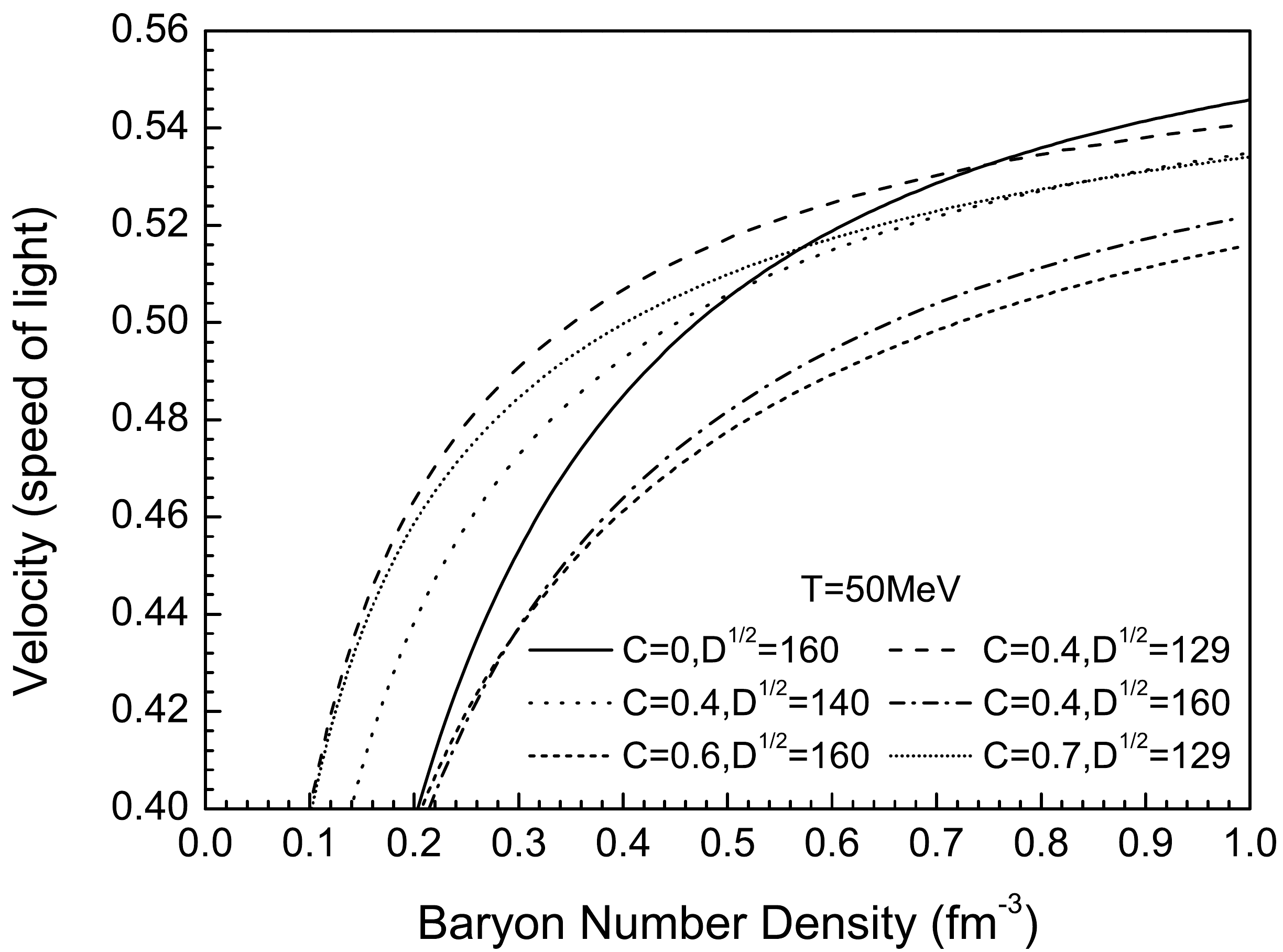}
\caption{The relation between sound velocity and baryon number in strange quark matter at $T=50\ \mathrm{MeV}$.}
\label{Fig5}
\end{figure}

Based on EOS indicated in Fig.~\ref{Fig4} and the sound velocity formula
\begin{equation}
v=\sqrt{\frac{\mathrm{d}P}{\mathrm{d}E}},
\end{equation}
we calculate the sound velocity in a SQM at fixed baryon number density.
As shown in Fig.~\ref{Fig5}, the sound velocity increases with the baryon number density.
In the case of high density, the curve corresponding to the larger $C$ and $D$ approximates the hyperelastic ($1/\sqrt{3}$) more slowly.
This is understandable because perturbation interactions still exist at higher densities.

\begin{figure}
\centering
\includegraphics[width=8cm]{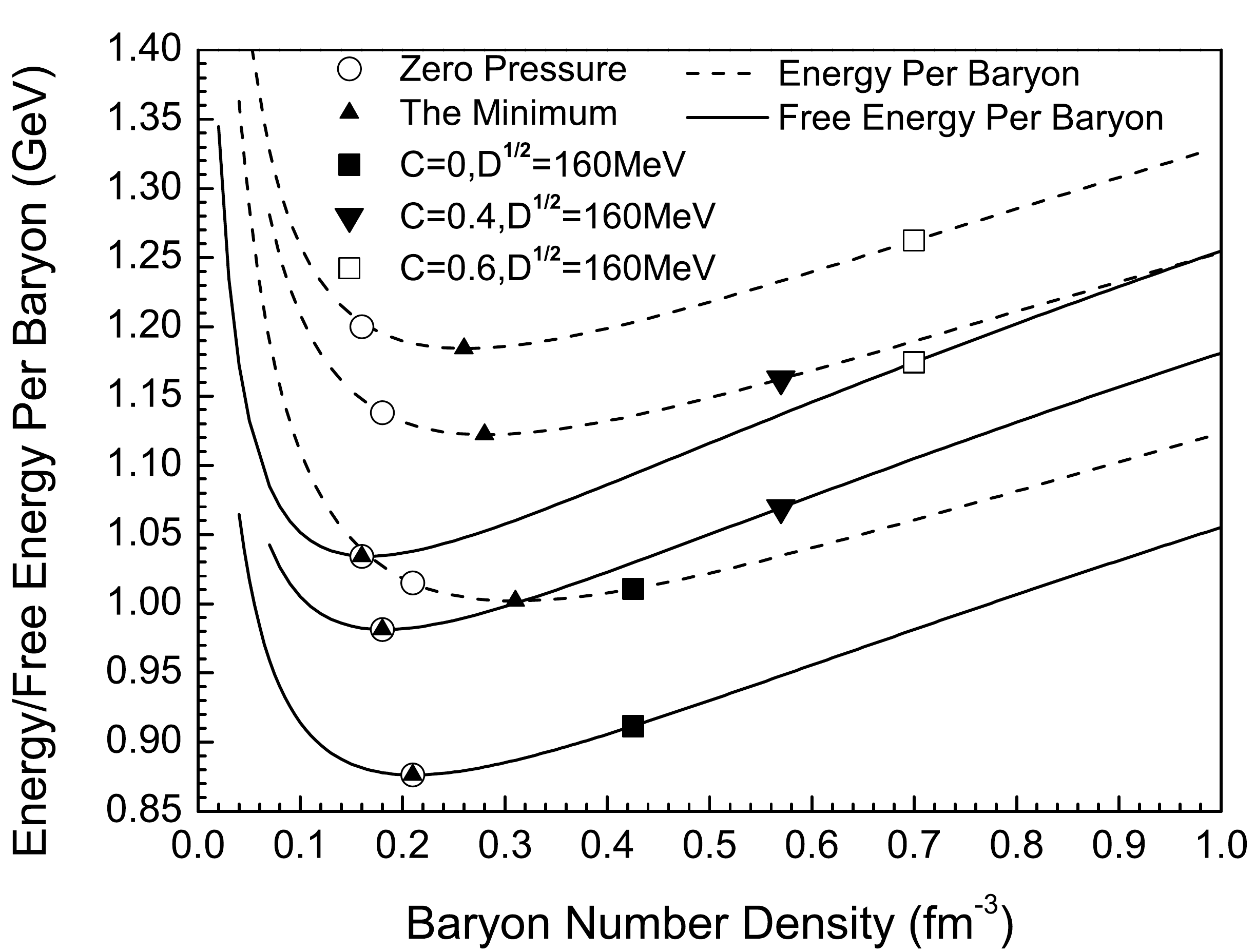}
\caption{The relationship between the free energy per baryon, energy per baryon and density at $T=30\ \mathrm{MeV}$. The triangle marks the minimum, and the open circles marks the zero pressure point.}
\label{Fig6}
\end{figure}

In order to check thermodynamic consistency and prove the influence of perturbation interaction, we plot the relationship between the free energy per baryon and energy per baryon and density at $T=30\ \mathrm{MeV}$ in Fig.~\ref{Fig6}, where $D^{1/2}=160\ \mathrm{MeV}$, $C=0$, $C=0.4$ and $C=0.6$.
The triangle is the minimum, and the open circles correspond to the zero pressure.
We notice that the pressure happens to be zero in the case of the minimum free energy, which is a necessary condition for thermodynamic consistency.

\begin{figure}
\centering
\includegraphics[width=8cm]{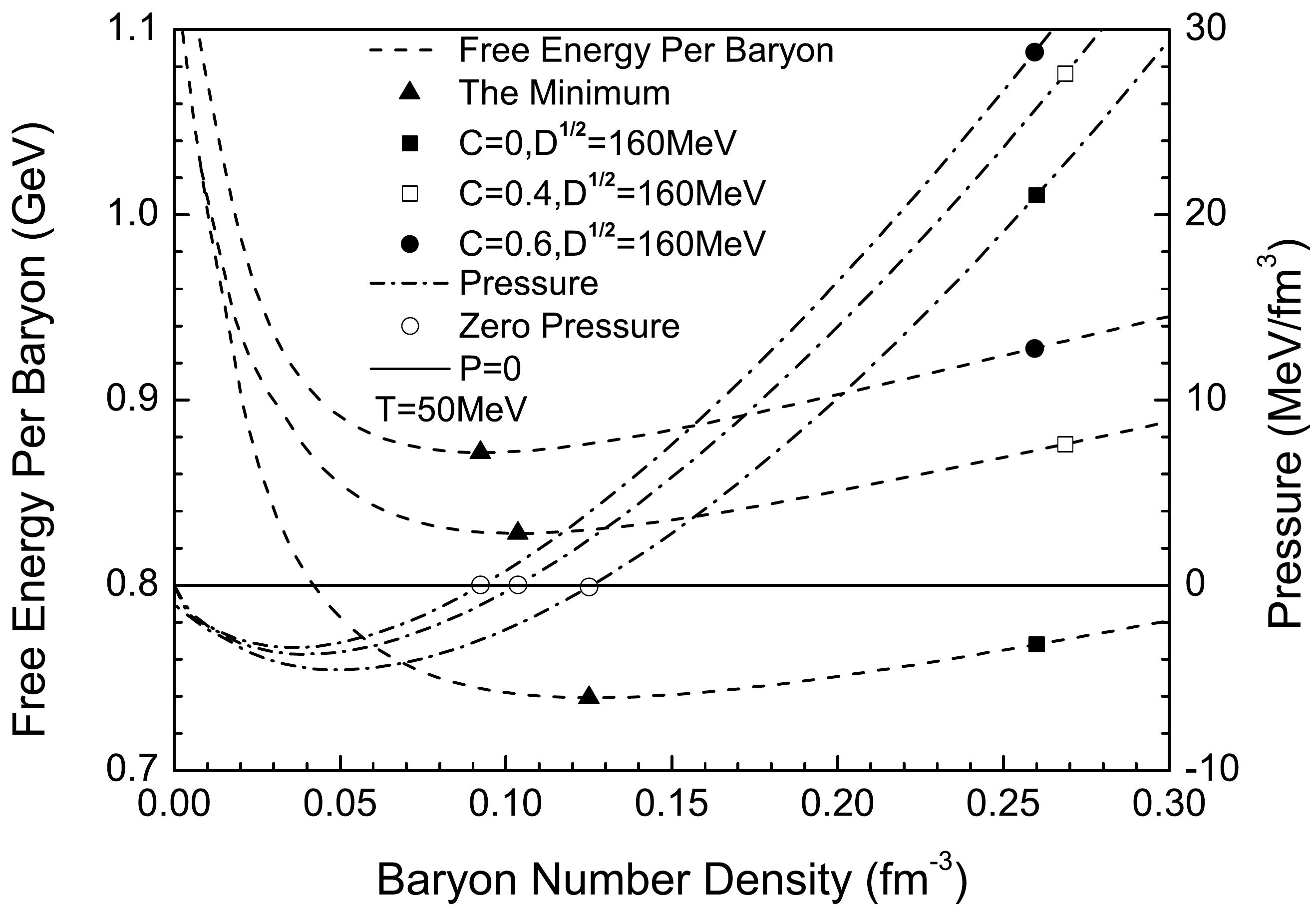}
\caption{The relationship between the free energy per baryon, pressure and density at $50\ \mathrm{MeV}$, where $D^{1/ 2}=160\ \mathrm{MeV}$, $C=0$, $C=0.4$ and $C=0.6$.}
\label{Fig7}
\end{figure}

Additionally, Figure~\ref{Fig7} shows the pressure and corresponding the free energy per baryon at $T=50\ \mathrm{MeV}$.
Obviously, when the density is less than the density with the lowest free energy, the pressure is negative.
Otherwise it's going to be positive.

To ensure the correctness of the results, we usually choose
\begin{equation}
P-n^{2}\frac{\mathrm{d}}{\mathrm{d}n}\left(\frac{F}{n}\right)=0,
\end{equation}
to test thermodynamic self-consistency. Please refer to Ref.\cite{Peng2000} for details of derivation.

Based on this formula, we can directly differentiate the average baryon free energy with respect to the baryon number density to get the pressure.
We note that the pressure at the lowest point of the energy per baryon is equal to zero, which is consistent with the most commonly used thermodynamic self-consistency test method.

\begin{figure}
\centering
\includegraphics[width=8cm]{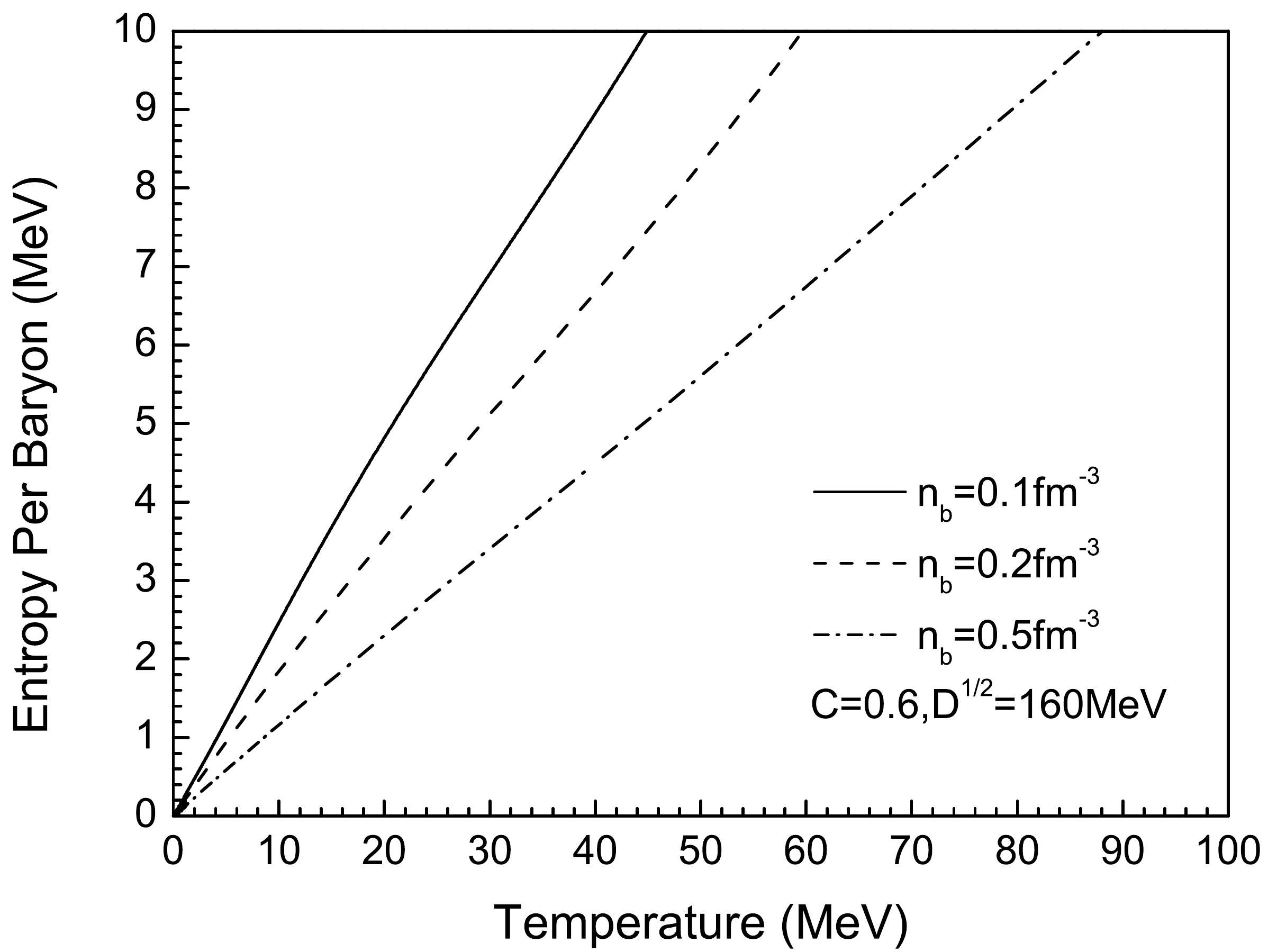}
\caption{The entropy per baryon is a function of temperature at different baryon number densities increases with temperature, where $D^{1/ 2}=160\ \mathrm{MeV}$ and $C=0.6$.}
\label{Fig8}
\end{figure}

Figure~\ref{Fig8} shows the entropy per baryon as a function of temperature at different densities.
It's an increasing function of temperature, whether the density is high or low, going to zero at zero temperature.
This is guaranteed by
\begin{equation}
\lim_{T\rightarrow 0}\partial m_{q} /\partial T=0.
\end{equation}

Interestingly, we did not require this in derivation of the partial derivative of mass with respect to temperature.
This ensures that at zero temperature, all quantities are restore to density-dependent models.
And of particular importance, as the temperature approaches zero, entropy naturally approaches zero, satisfying the third law of thermodynamics.

\section{properties of strange stars}
\label{stars}

Quark stars are an interesting branch of nuclear physics, astrophysics, and other important fields. In the past two decades,  there has been a lot of studies on this matter, such as Refs.\cite{Alford2007,Weissenborn2011,Dexheimer2013,Klaehn2013,Deb2018}.

\begin{figure}
\centering
\includegraphics[width=8cm]{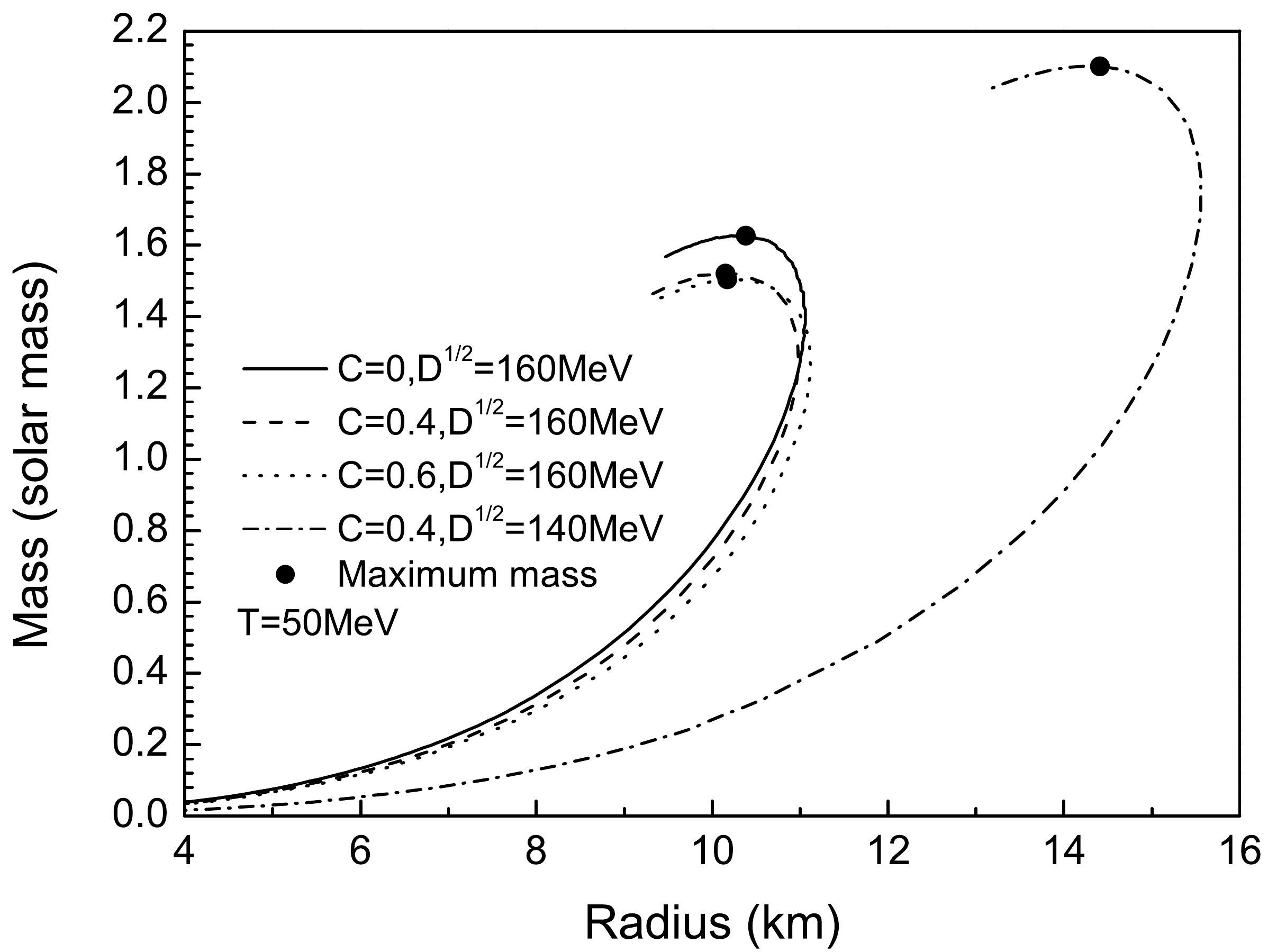}
\caption{The relationship between the mass and radius of the strange star corresponding to different perturbation parameters at $50\ \mathrm{MeV}$, where $D^{1/ 2}=160\ \mathrm{MeV}$, $C= 0$, $C= 0.4$ and $C= 0.6$ and $D^{1/ 2}=140\ \mathrm{MeV}$, $C= 0.4$.}
\label{Fig9}
\end{figure}

Based on the EOSs indicated in Fig. 4, we solve the Tolman-Oppenheimer-Volkov equation
\begin{equation}
\frac{\mathrm{d}P}{\mathrm{d}r}=-\frac{GmE}{r^{2}}\frac{(1+P/E)(1+4\pi r^{3}P/m)}{1-2Gm/r},
\end{equation}
and

\begin{equation}
\mathrm{d}m=4\pi Er^{2}\mathrm{d}r.
\end{equation}
Then the masses and radii of the strange star corresponding to different parameters are obtained.
The results are shown in Fig.~\ref{Fig9}, where the most massive stars are represented by black dots.
Generally speaking, the maximum star mass increases with the perturbative strength $C$ and decreases with confinement intensity parameter $D$.
We can see that relative to the previous quark mass scaling, the maximum star mass exceeds $2M_{\odot}$ after considering the first-order perturbation interaction.

\begin{figure}
\centering
\includegraphics[width=8cm]{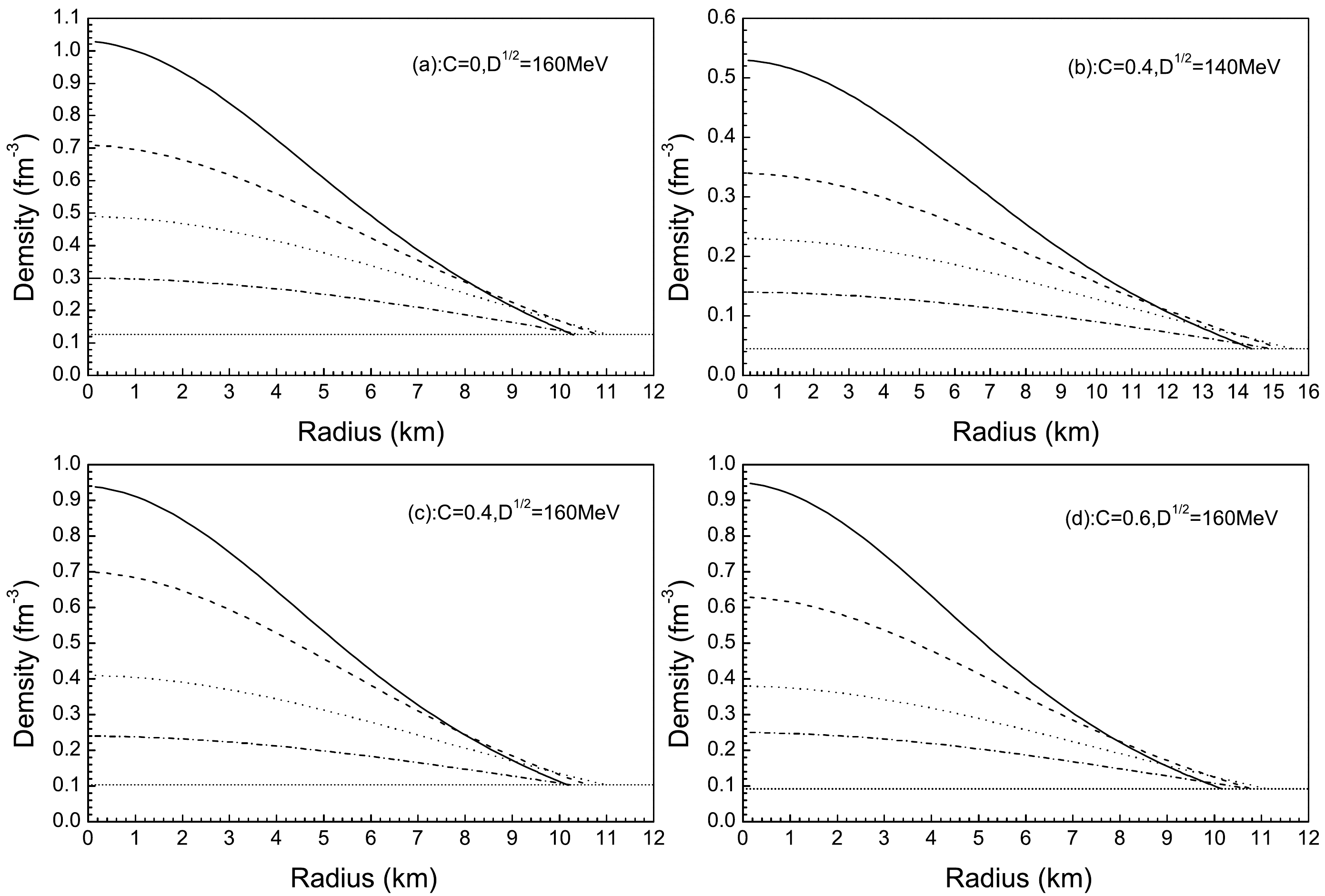}
\caption{Density distribution of different parameter groups at $50\ \mathrm{MeV}$. The top curve of each subgraph is the central density corresponding to the maximum mass of the star, the third curve is the central density corresponding to the maximum radius of the star, and the horizontal line corresponds to the surface density of the star.}
\label{Fig10}
\end{figure}

We found that the baryon number density on the surface of the star would become very small, or even lower than the normal nuclear saturated density when larger perturbation intensity parameters $C$ and smaller quark confinement parameters $D$ were used to generate the maximum mass of a large star.
This is exactly the signal of phase transition of nuclear matter, so in the study of quark mass scale, it is necessary to further study QCD phase diagram.
In order to clearly illustrate this point, we chose four sets of parameters and drew the radial distribution diagram of the density corresponding to  the center baryon number density  of  different stars, as shown in the  Fig.~\ref{Fig10}.

We can see, the surface density of the star becomes very small, even lower than the normal nuclear saturation density when a larger $C$  is used to produce a larger maximum mass.
The panel (a) is for the parameter pair $(C,\sqrt{D})=(0,160\ \mathrm{MeV})$.
At this point, the density of baryon number corresponding to the maximum star mass (about 1.63 $M_\odot$) is 1.03 $\mathrm{fm}^{-3}$.
The surface density is 0.126 $\mathrm{fm}^{-3}$.
When the center density of baryon number is 0.5 $\mathrm{fm}^{-3}$, the maximum radius of the star is about 11.5 km.
The panel (b) is for the parameter pair  $(C,\sqrt{D})=(0.4,140\ \mathrm{MeV})$.
At this point, the density of baryon number corresponding to the maximum star mass (about 2.10 $M_\odot$) is 0.53 $\mathrm{fm}^{-3}$.
The surface density is 0.045 $\mathrm{fm}^{-3}$.
When the center density of baryon number is 0.23 $\mathrm{fm}^{-3}$, the maximum radius of the star is about 15.56 km.
The panel (c) is for the parameter pair $(C,\sqrt{D})=(0.4,160\ \mathrm{MeV})$.
At this point, the density of baryon number corresponding to the maximum star mass (about 1.51 $M_\odot$) is 0.975 $\mathrm{fm}^{-3}$.
The surface density is 0.103 $\mathrm{fm}^{-3}$.
When the center density of baryon number is 0.41 $\mathrm{fm}^{-3}$, the maximum radius of the star is about 10.98 km.
The panel (d) is for the parameter pair $(C,\sqrt{D})=(0.6,160\ \mathrm{MeV})$.
At this point, the density of baryon number corresponding to the maximum star mass (about 1.50 $M_\odot$) is 0.95 $\mathrm{fm}^{-3}$.
The surface density is 0.093 $\mathrm{fm}^{-3}$.
When the center density of baryon number is 0.38 $\mathrm{fm}^{-3}$, the maximum radius of the star is about 11.12 km.

In these cases, the maximum mass of the star increases, but the surface density decreases, even below normal nuclear matter.
This is obviously a signal of a phase transition to nuclear matter.
Therefore, further research on QCD phase diagram is necessary for the follow-up work of the new quark mass scaling method.

\section{SUMMARY}
\label{sec:sum}

The paper studied the thermodynamic self-consistency of the equivalent mass model in detail.
Combined with new quark mass scale,  we studied the equation of state of strange quark matter at finite density and temperature.
Next, we study the effects of quark confinement interaction and first order perturbation interaction on strange quark matter.
Based on the equation of state, we calculate the velocity of sound of strange quark matter, the relationship between  strange star and radius, where include the maximum stellar mass varies with perturbation intensity parameters and constraint intensity parameters and the relationship between the mass and radius of strange star whose surface density is lower than that of normal nuclear matter.
We think this could be a signal for a phase transition to nuclear matter.
The self-consistency of thermodynamic treatment of the model is verified by the relationship between pressure and free energy.
We have verified the third law of thermodynamics for strange quark matter under this model by the entropy of the average baryon as a function of temperature.

Next, we plan to study the phase diagram of the strange quark matter under the equivalent mass model and compare it with the phase diagram of the strange quark matter under the traditional MIT model.
Then we will study the application of the phase diagram, where include the structure of hybrid stars, the properties of strangelets,  extreme relativistic heavy ion collisions, etc.

\appendix

\section*{ACKNOWLEDGMENTS}

The authors would like to thank the support from National Natural Science Foundation
of China (Grant No. 11875052).

\end{document}